# Review of existing breakup and transfer theories leading to continuum final states[1]

Angela Bonaccorso

Institute for Nuclear Theory, Univ. of Washington, Seattle, WA 98915, USA,

Istituto Nazionale di Fisica Nucleare, Sez. di Pisa, 56100 Pisa, Italy,[2]

Bonaccorso@pi.infn.it

**Abstract**

This is a brief review of recent experimental and theoretical works aiming at extracting spectroscopic information from quasielasic projectile breakup reactions at intermediate energies.

## 1  Introduction

In this paper I will restrict myself to the study of one neutron nuclear breakup from single particle orbitals in quasi elastic reactions. The motivation for such a choice is that recently for this kind of reactions the emphasis has been shifting from the understanding of the reaction mechanism to the extraction of structure information on the reaction partners and some remarkable progress have been made. In the past several reaction models have been successful in describing projectile breakup into various types of fragments but we believe that one nucleon breakup is so far the best example of how one can go beyond the reaction mechanism to the extraction of structure information. I will discuss two specific situations: one deals with very heavy ions like $^{20}Ne +^{208} Pb$, the other with lighter nuclei like $^{11}Be +^9 Be$ or $^{12}Be +^9 Be$ in which the projectile is a halo nucleus like $^{11}Be$ or simply a weakly bound nucleus like $^{12}Be$. In the first example the structure information extracted concerns the position and widths of the single particle resonances in the continuum of $^{208}Pb$ and the extraction of the three-body physical background. Also it has been shown that inclusive spectra obtained from measurements of the projectile energy loss contain contributions from several initial states in the projectile core which lead to a slightly excited ejectile. This happens when the excitation energy is within the experimental energy resolution. This characteristic of breakup reactions has been the origin of interesting experimental studies for the second category of nuclei mentioned above, namely halo or weakly bound nuclei. In this case in fact a simple measurement of the $\gamma$-rays in coincidence with the core has allowed for detailed spectroscopic studies of the projectile nucleus. In all cases the incident energies per nucleon are in the range $40 - 80 A.MeV$, thus allowing for simple semiclassical approximations to be introduced.

---

[1] Invited talk given at the 9th Internatinal Conference on Nuclear Reaction Mechanisms, Varenna, Italy, 2000

[2] Permanent address

I will describe a semiclassical model called transfer to the continuum (TC) introduced by Bonaccorso and Brink in 1988 and I will show how it can be considered as an intermediate model between fully quantum mechanical theories like DWBA and high energy eikonal models.

## 2 Semiclassical models

The interest of the TC model lies in the fact that it can be obtained from the formula for the DWBA amplitude if a semiclassical approximation is made on the ion-ion distorted waves. The rest of the amplitude, dealing with the initial and final neutron interaction with target and projectile core respectively, is treated fully quantum mechanically. On the other hand it can be shown that in the case of a very weak initial binding the formulae have a simple eikonal limit. Therefore the TC model seems to be a useful intermediate alternative between a complete quantum mechanical treatment of the reaction mechanism and an extreme classical limit of it. A key point of the model is the use of asymptotic wave functions for the initial and final neutron states. In this way simple analytical formulae can be obtained for the breakup probability. Unfortunately this approximation restricts the applicability of the method to the part of the reaction corresponding to large core target impact parameters for which the basic hypothesis of no-overlap between the neutron initial and final potentials is valid.

The starting point for the derivation of the TC model was D. M. Brink semiclassical model for transfer reactions. Our original motivation was to try to get some insight into the physical content of full coupled channels (CC) and/or DWBA methods as described for example in [1]-[6] and eikonal models like [7]-[9].

The semiclassical method of Brink and collaborators starts from the hypothesis that the DWBA T-matrix can be simplified [10, 11] if one uses a WKB approximation to the distorted waves describing the relative ion-ion motion in the initial and final channels. Similar hypothesis were made by the Copenhagen group [12] or by other authors [7, 13]. Within the approximation of a classical trajectory it is clear that the ambiguity present in the Baur and al. method [3] or in Udagawa and Tamura method [5], whether it would be better to use the post or prior form of DWBA, is resolved since the incoming and outgoing channel trajectories are equal.

Under the approximation that the ion-ion relative motion trajectory can be taken as a straight line with constant velocity $R(t) = b_c + vt$ and supposing strong core-target absorption and no overlap between the core and target binding potentials such that the only dynamic role of core-target wave function is to conserve energy, one obtains for the differential breakup cross section [14, 15]

$$\sigma_{bup}(\xi) = C^2 S \int_0^\infty d^2 b_c \ P_{bup}(b_c, \xi) \ |S_{ct}(b_c)|^2, \quad (1)$$

where $b_c$ is the core-target impact parameter in the xy-plane perpendicular to the relative motion velocity v taken along the z-axis. $|S_{ct}|^2$ is the core survival probability, $\xi$ is any differential variable like energy or momentum with respect to which experimental measurements can be made. $P_{bup}$ can in principle be obtained from an exact transition amplitude as discussed in [16]:
$A = \lim < \Phi_{f,out}(t_2)|G(t_2,t_1)|\Phi_{i,in}(t_1) >$, where G is the full propagator taking into account final state interactions of the neutron with both core and target potential, and $\Phi_{in,out}$ are asymptotic states in the initial and final channel, and the limit is taken for $t_2 \to \infty, t_1 \to -\infty$. If the most important final state interaction is with the target then the breakup amplitude reduces to

$$\begin{aligned} A_{bup} &= \frac{1}{i\hbar} \int_{-\infty}^\infty dt < \psi_2(t)|V_{nt}(r)|\psi_1(t) > \\ &\approx \int dk_y \sqrt{k_y^2 + \eta^2} \bar{\psi}_i(d_1, k_y, k_1) \bar{\psi}_f^*(d_2, k_y, k_2) \end{aligned} \quad (2)$$

Eq.(2) shows explicitly that the transfer or breakup amplitude is given by an overlap of the initial and final neutron momentum distributions and $d_1$ and $d_2$ are the neutron positions with respect to the core and target when the two nuclei are at the distance of closest approach. The initial and final momenta along the z-direction are fixed by the kinematic and their values are:
$k_1 = (-\varepsilon_i + \varepsilon_f - \frac{1}{2}mv^2)/\hbar v \quad k_2 = (-\varepsilon_i + \varepsilon_f + \frac{1}{2}mv^2)/\hbar v$ where $\varepsilon_i$ and $\varepsilon_f$ are the initial and final bound and continuum neutron energies respectively. The transverse neutron momentum is conserved and it is given by $\eta^2 = \gamma_i^2 + k_1^2 = k_2^2 - k_f^2$ with $\gamma_i^2 = -2m\varepsilon_i/\hbar^2 \quad k_f^2 = 2m\varepsilon_f/\hbar^2$ so that the best matching condition leading to a peak in the energy or parallel momentum spectrum of the neutron is obtained when the classical energy conservation condition is satisfied $\varepsilon_f = \varepsilon_i + \frac{1}{2}mv^2$. $\frac{1}{2}mv^2$ is the incident energy per nucleon at the distance of closest approach.

The hypothesis of non-overlapping potentials reminds of Bég theorem [17] which in turn is a justification for the so-called surface approximation [18]. Thanks to this approximation one is allowed to use Hankel forms for both the initial and final state leading to an asymptotic single particle wave function for the initial state :

$$\psi_i(\mathbf{r},t) = C_i \ \gamma_i \ h_{l_i}^{(+)}(i\gamma_i r) Y_{l_i m_i}(\Omega_i) e^{-\frac{i}{\hbar}\varepsilon_i t} \qquad (3)$$

On the other hand two possible approximations for the final state wave functions are:

a quantum mechanical one

$$\psi_f(\mathbf{r},t) = -C_f \frac{i}{2} k_f \left( h_{l_f}^{(-)}(k_f r) - S_{l_f} h_{l_f}^{(+)}(k_f r) \right) Y_{l_f m_f}(\Omega_f) e^{-\frac{i}{\hbar}\varepsilon_f t} \qquad (4)$$

and its eikonal limit

$$\psi_f(\mathbf{r},t) = e^{i\mathbf{k}_\perp \cdot \mathbf{b}} e^{ik_z z} e^{\frac{i}{\hbar v} \int_z^\infty V_{nt}(x,y,z')dz'} e^{-\frac{i}{\hbar}\varepsilon_f t} \qquad (5)$$

Finally the differential cross section with respect to the neutron transverse momentum, including spin coupling [19, 20] is

$$\frac{dP_{bup}}{dk_1} \approx \frac{1}{2}\Sigma_{j_f}(2j_f+1)(|1-\langle S_{j_f}\rangle|^2 + 1 - |\langle S_{j_f}\rangle|^2)(1+R)B_{l_f,l_i}. \qquad (6)$$

$$B_{l_f,l_i} = \left[\frac{\hbar}{mv}\right]\frac{1}{k_f}|C_i|^2 \frac{e^{-2\eta d}}{2\eta d} P_{l_i}(X_i) P_{l_f}(X_f), \qquad (7)$$

where $X_i = 1 + 2(k_1/\gamma_i)^2$   $X_f = 2(k_2/k_f)^2 - 1$. The two terms in Eq.(6) come from the average over the neutron continuum energy of the factor $|1 - S_j|^2$, this factor in turn comes from the fact that the hankel functions $h_{l_f}^+$ and $h_{l_f}^-$ in the final continuum wave function have the same Fourier transform. If we calculate the energy averaged neutron target S-matrix $<S_j>$ by using an appropriate energy dependent optical potential, then it is clear that with the term $1 - |<S_j>|^2$ we shall be able to describe consistently single particle isolated resonances, as done by the Udagawa and Tamura method [5], strongly overlapping resonances giving rise to compound nucleus as well as smooth resonances and also to the so called inelastic breakup of Baur et al.[3]. All other averaged fluctuations are described, the present approach, by the term $|1-<S_j>|^2$ which represents part the physical three-body background due simply to the elastic scattering of the breakup neutron by the target.

The above partial wave expansion in terms of the neutron-target angular momentum, reminds of similar expansions present in projectile spectator models as reviewed for example by Mc Voy[2]. However it is interesting to notice that in our case, because of the overlapping momentum distributions Eq.(2) and of the accurate definitions of the neutron momenta $k_1$ and $k_2$, which take into account neutron energy and momentum conservation there is no complete factorization into a neutron-target cross section and an initial momentum distribution. This point can be further clarified by looking at the eikonal limit of Eq.(6). It can be obtained by making a simple correspondence between the sum over partial waves with the integral over the neutron impact parameter with respect to the target [21]. Otherwise it can be obtained directly from the amplitude Eq.(2) by using the final eikonal wave function Eq.(5) as outlined in [22]

$$\frac{dP_{bup}}{dk_1} \approx \int_0^\infty d^2\mathbf{b_n} \left(|1 - e^{-i\chi(b_n)}|^2| + 1 - e^{2Im\chi(b_n)}\right) \bar{\psi}_i(|\mathbf{b_n} - \mathbf{b_c}|, k_1)|^2 \quad (8)$$

The above expression contains explicitly the overlap between the initial momentum distribution of the neutron and its profile function with respect to the target. In the limit of a vanishing initial binding energy the dependence on $b_c - b_n$ is very slow and can be approximated with $b_c$ so that it is a reasonable approximation to take the momentum distribution outside the integral. In this way the above formula looks similar to the old one of the PWBA [4]. Also in this limit if the eikonal S-matrix in the diffraction probability integral, first term of Eq.(8) is expanded in powers of the phase shift $\chi(\mathbf{b}) = \frac{1}{\hbar v} \int_{-\infty}^\infty e^{i(k_2-k_z)z'} V_{nt}(x,y,z')dz' \approx \frac{1}{\hbar v} \bar{V}_{nt}(\mathbf{b}, 0)$ and only the first order term is retained, we are left with the very well known Serber formula. Here as in all previous equations where it appears, $V_{nt}(\varepsilon_f)$ is an energy dependent n-target optical potential of finite range.

Having clarified the reaction mechanism we proceed now to the discussion of the structure information we can extract from data.

## 3 Comparison of theoretical and experimental results

### 3.1 Heavy nuclei: the role of the target

I will start this section by discussing Fig.(1) where a series of experimental spectra and calculations are collected [23]-[28]. The data represent target excitation energy spectra obtained from the ejectile energy loss spectra. Our corresponding TC calculations are obtained from the neutron spectra using energy conservation. Some spectra show explicitly the calculated elastic breakup contribution to the physical background. It comes from the first term in the RHS of Eq.(6). This part gives always about one third of the total cross section. The inelastic breakup and absorption into single particle resonances are both contained in the second term of Eq.(6).

For heavy projectiles the initial separation energies are around 15MeV. This means that for a typical incident energy of 40A.MeV the neutron most favorite energy in the continuum will be around 25MeV. At this energy high angular momentum single particle resonances in the continuum are still not completely dumped[26] and they show up in Fig.(1) from [25] superimposed to the physical background.

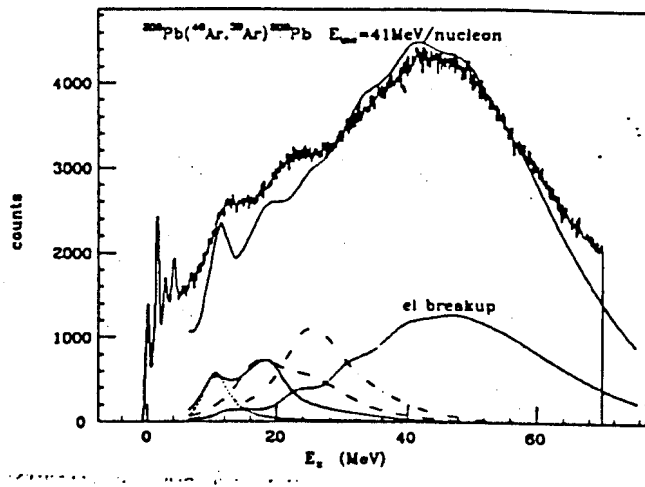

FIG. 1a. Inclusive spectrum of the reaction $^{208}$Pb($^{40}$Ar,$^{39}$Ar)$^{209}$Pb at $E_{inc}$ = 41 MeV/nucleon. The solid curve superimposed onto the experimental spectrum is the result of our calculation for the cross section due to transfer from the $1d_{3/2}$ initial state in Ar. In the lower part of the figure the dotted curve shows the contribution of the $1k_{17/2}$ final state. The solid curve is the total contribution due to $l_f$ = 8. The second peak is due to the $1k_{15/2}$ state. The dashed line is the contribution of $l_f$ = 9 and the dot-dashed line is for $l_f$ = 10. The tightly dotted curve is the elastic breakup.

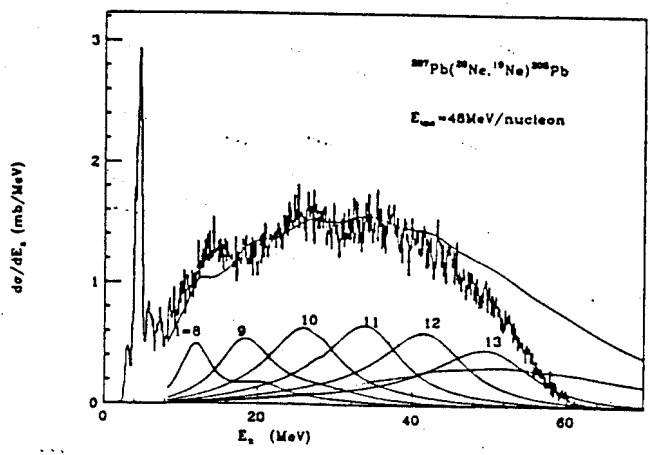

FIG. 1b. Inclusive spectrum of the reaction $^{207}$Pb($^{20}$Ne,$^{19}$Ne)$^{208}$Pb at $E_{inc}$ = 48 MeV/nucleon. The solid curve superimposed onto the experimental spectrum is the result of our calculation. The dotted curve shows the calculated breakup while the curves labeled $l_f$ = 8–13 show individual contributions to the absorption cross section of several final angular momenta.

In order to extract the single particle state strength it is necessary to make coincidence experiments [27] between the ejectile and neutron and at the same time reconstruct the final excitation state of the target. Examples of how the inelastic breakup has been studied can be found in [3, 23, 28]. For incident energies around 60A.MeV or higher elastic and inelastic breakup are always dominant [24]. At this point I would like to mention that all the calculations shown in Fig.(1) were possible thanks to the use of the excellent neutron-$^{208}$Pb optical potential parameterization of Mahaux and Sartor [29]. In fact the reproduction of the proper relative amount of elastic vs. absorption cross section and the positions and widths of the single particle resonances are only possible if Eq.(6) is calculated with a S-matrix obtained from a different potential at each neutron final energy in the continuum. Such S-matrices must in turn give the correct elastic and reaction free neutron-target cross sections at each energy (cf. Fig.(4)). Finally we show in Fig.(2) the initial state momentum distributions for different angular momenta. The measured spectra of Fig.(1) look quite different from them but we shall see how this feature changes when the the neutron in the initial state is weakly bound.

## 3.2 Halo and weakly bound nuclei: the role of the projectile

A number of nuclear breakup experiments [30]-[42] have been performed at various laboratories to study weakly bound light nuclei and search for halo structures. $^{11}$Be and $^{11}$Li have been clearly identified as halo nuclei [30, 31, 32]. $^{19}$C is the heaviest known nucleus with a close-to-a-halo structure, having a ground state wave function which contains 60% of $2s_{1/2}$ state with a separation energy of around 0.5MeV [36, 37, 42, 43, 20]. Other "exotic" nuclei exist and are interesting to study as probe of nuclear models. For example the breakdown of the N=8 shell closure in $^{12}$Be has been recently demonstrated experimentally [41]. Theoretically one knows that deformation and paring effects as described for example in [44] are responsible for the appearance of intruder s-states and consequent modification of the shell model level sequence. A more precise determination of the single particle structure of light, weakly bound projectiles, is therefore now possible, since the reaction mechanism is well understood and described by models like the TC or its eikonal limit. In fact in the case of a weakly bound projectile the best matching condition for the neutron final energy discussed in Sec.2 gives that the final neutron energy is very close to the incident energy per nucleon. Since the incident energies are quite high, this means that the neutron final energy with respect to the target is also large, typically around 40MeV or higher. At this energy no target exhibits single particle resonances any more and the breakup spectra are dominated by elastic and inelastic breakup. They are therefore rather featureless and will not give any information on the target structure. On the other hand since the eikonal limit Eq.(8) becomes more valid, it is easy to understand that these spectra show very close similarity to the neutron initial momentum distribution as shown in Fig.(2) although their widths will be somewhat modified by the dynamics of the interaction with the target as described by the neutron-target and core-target S-matrices and explained in [23], [45]-[50]. Thus they have been used to identify the initial angular momentum state of the neutron in the projectile. In Fig.(3) we show the experimental [41] and calculated [47] spectra. Eq.(6) gives similar spectra, a part from the absolute normalization, if the spin-orbit coupling is neglected. For initial states with $l_i > 1$ we found [20] that the effect of the spin-orbit is to give rather asymmetric spectra. It would be interesting to see if future experiments will confirm such a trend.

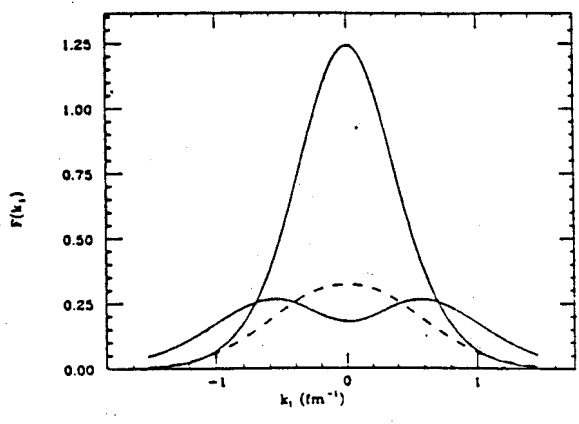

FIG. 2 Initial-state momentum distributions in $^{20}$Ne according to Eq. (2.3a). The solid curve is for the $2s_{1/2}$ state, the dashed curve is the for $1p_{1/2}$, while the dotted curve is for the $1d_{5/2}$ state.

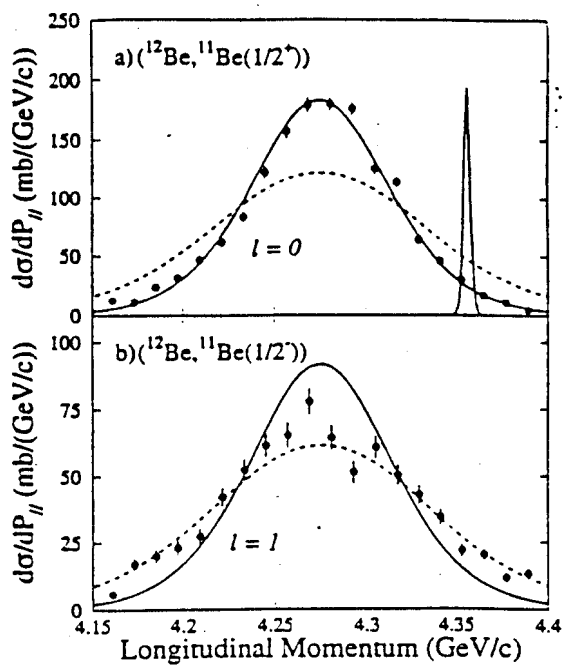

Fig. 3

Finally absolute cross sections can be obtained with both the TC and eikonal formalism within 20% accuracy [51], thus allowing for the extraction of spectroscopic factors. This method has been used by the authors of refs.[38]-[42] to interpret their experiments in which $\gamma$-rays in coincidence with the core have been detected to identify the initial state from which the breakup neutron de-cays. Comparison between the measured cross sections and calculations made with the eikonal model of [52] as used in [53] allows to extract the experimental spectroscopic factors which are then compared to shell model values [54].

As we mentioned in the previous section an essential ingredient of the calculations with both the TC method and with the eikonal model, is the optical potential used to calculate the neutron-target S-matrix. In the case of nuclear breakup experiments discussed here a $^9$Be target has very often been used. Since there is no phenomenological potential available for n-$^9$Be scattering in the energy range 20-180MeV, we have extended and modified the parameterization given in [55]. We show in fig.(4) from [51] the calculated total cross sections (full curve) together with the experimental points from [56] and elastic and reaction calculated cross sections (dotdashed and dashed lines respectively). For a microscopic optical potential we refer the reader to Prof. Amos's talk in these proceedings and also to the contribution of Dr. Chadwick for more information on neutron-light nuclei cross sections. Using our potential we have made some comparison between the absolute cross sections obtained from the TC model and the eikonal model. We have found a good agreement between the two models, the eikonal giving typically cross sections larger by 20%. This is partially due to the fact that the eikonal model does not conserve energy [8], but the main origin of the difference comes from the fact that for such light target diffraction and reflection quantum mechanical effects are quite important and they are not reproduced by the eikonal model because of the straight line trajectory approximation for the neutron-target relative motion. Our calculated values together with the results of ref. [41] are given in table 1.

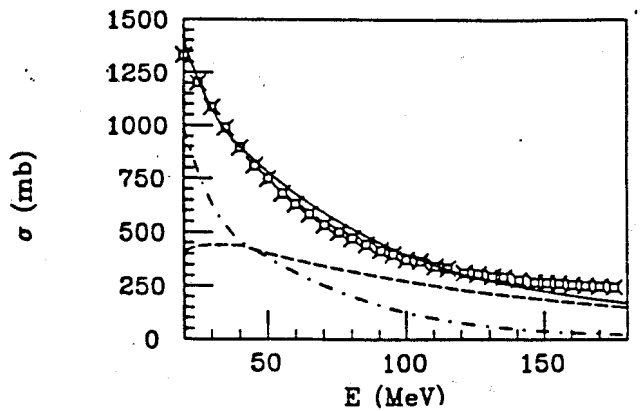

Fig.(4) . Neutron-$^9Be$ calculated and experimental cross sections.

I would like to thank Prof P.G. Hansen and Co. for communicating their results prior to publication.Table 1. Calculated [51] breakup cross sections in mb for $^{12}Be$ on $^9Be$, at 78 A.MeV [41]. The first row contains values for breakup from the $2s_{1/2}$ orbital in $^{12}Be$, the second row of the table is for breakup from the $1p_{1/2}$ orbital, at several incident energies. Energies in A.MeV. $S_{SM}$ are the spectroscopic factors quoted in [41].

| $E_{inc}$ | $\sigma_{str}$ TC | $\sigma_{str}$ eik | $\sigma_{diff}$ TC | $\sigma_{diff}$ eik | $\sigma_{-n}$ TC | $\sigma_{-n}$ eik | $\sigma_{exp}$ | $S_{SM}$ | $S_{TC}$ | $S_{eik}$ |
|---|---|---|---|---|---|---|---|---|---|---|
| 78 | 43 | 52 | 19 | 24 | 62 | 76 | 32(5) | 0.69 | 0.65 | 0.53 |
| 40 | 23.4 | 29.6 | 16.3 | 18 | 39.7 | 47.6 | | | | |
| 60 | 27.2 | 32.7 | 13 | 15.9 | 40.2 | 48.6 | | | | |
| 78 | 28 | 34 | 11 | 13 | 39 | 47 | 18(3) | 0.58 | 0.54 | 0.45 |
| 100 | 28.4 | 33.7 | 8.4 | 9.7 | 36.8 | 43.4 | | | | |

# References

[1] A. K. Kerman and K. W. McVoy, Ann. Phys. **122** 197 (1979).

[2] K. W. McVoy, Proceedings of the workshop on 'Coincident Particle Emission from Continuum States in Nuclei', Ed. H.Machner and P. Janh, World Scientific, 1984.

[3] G. Baur, F. Rosel, D. Trautmann, and R. Shyam, Phys. Rep. **111** 333 (1984) and references therein.

[4] R. J. de Meijer and R. Kamermans, Rev. Mod. Phys.,**57** 147 (1985).

[5] T.Udagawa and T. Tamura, Phys. Rev. C **24** 1384 (1981).

[6] N. Austern C. M. Vincent, Phys. Rev. **C23** 2847 (1981).

[7] M.S.Hussein ,K.W. McVoy and D. Saloner , Phys. Lett. **B98** 162 (1981).
M. S. Hussein and K. W. McVoy, Nucl. Phys. **A445** 124 (1985).

[8] T. Fujita and J. Hüfner, Nucl. Phys. **A** 493 (1980).

[9] J. Hüfner and M. C. Nemes, Phys. Rev. **C23** 2538 (1981).

[10] H. Hasan and D. M. Brink, J. Phys. **G4** 1573 (1978).
10